# Influence of a Medium on Capacitive Power Transfer Capability

Lecluyse Cédric
*Department of Electrical Engineering (ESAT)*
*KU Leuven, ELECTA Ghent*
Ghent, Belgium
cedric.lecluyse@kuleuven.be

Minnaert Ben
*Department of Industrial Science and Technology*
*Odisee University College of Applied Sciences*
Ghent, Belgium
ben.minnaert@odisee.be

Ravyts Simon
*Department of Electrical Engineering (ESAT)*
*KU Leuven, ELECTA Ghent*
Ghent, Belgium
simon.ravyts@kuleuven.be

Kleemann Michael
*Department of Electrical Engineering (ESAT)*
*KU Leuven, ELECTA Ghent*
Ghent, Belgium
michael.kleemann@kuleuven.be

*Abstract*— Despite the advantages of capacitive power transfer (CPT), inductive power transfer (IPT) is still preferred. The reason: IPT systems have a gap power density in air that is 400 times greater. Conclusively, IPT can transmit more power than CPT over greater distances in air, but what about other media? This paper gives an answer on how media, different from air, influence the power transfer over different distances. First, we analyze theoretically the capacitive coupling with different media in the gap. Next, we simulate the CPT system using finite element software and compared it with the theoretical analysis. Finally, we employ the results of the finite element simulation in a power electronic simulation to examine the influence of the medium on the electrical power transfer.

*Keywords*— wireless power transmission, capacitive power transmission, capacitive coupling, electric coupling, electric fields.

## I. INTRODUCTION

Wireless power transfer (WPT) is an emerging technology that enables more convenient, standardised and safer systems [1]. The most common WPT technology is inductive power transfer (IPT). This technology is based on energy transfer with magnetic fields. It employs coils at the primary and secondary side. Another WPT technology is capacitive power transfer (CPT). Here, energy is transferred via electric fields between metal plates at primary and secondary side [1]. Compared to IPT, CPT has the advantages of being cost effective, it does not bear transmission losses in the vicinity of metal objects and emits fewer EM disturbances [1]–[3].

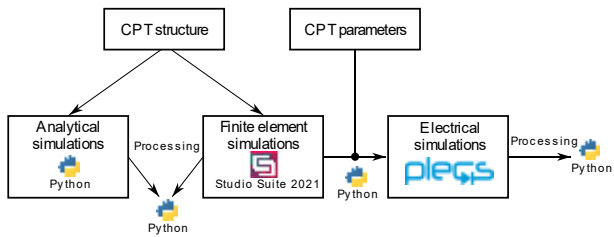
Fig. 1: Implemented methodology

An important performance criterion of WPT is gap power density. The gap power density of IPT is 400 times greater than CPT with air as a medium [4]. This results in IPT being able to transmit higher powers over longer distances. For this reason, IPT is generally preferred over CPT for energy transfer through air. However, considering other solid or liquid media in the gap can be in favour of CPT technology. Also, CPT can become feasible in applications where IPT is limited by core materials [4].

This paper will investigate the research question: How do media different from air influence the power transfer of a CPT system over different distances? The methodology to investigate this question is shown in Fig. 1. First, an analytical approach will determine the coupling and leakage capacitances with different media. Second, finite element simulations will validate the analytical findings and yield more detailed results. Third, the obtained capacitances for different media are employed in an electrical simulation of the CPT that takes limitations of present semiconductor switches into account. This simulation will yield the influence of the different media on the electrical power that can be transferred.

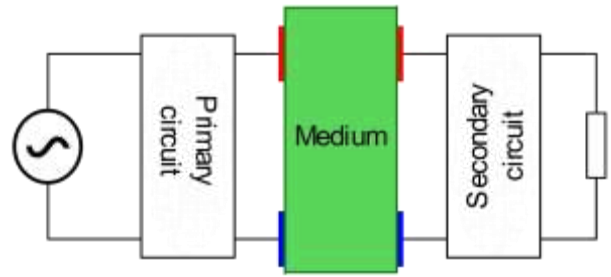
Fig. 2: Simple CPT structure

## II. CAPACITIVE COUPLING

A CPT system is commonly composed of two parallel pairs of metal plates with a medium between them, as shown in Fig. 2. A more detailed schematic equivalent circuit representation of this so-called four-plate structure is given in Fig. 3. The system consists of a pair main coupling capacitors ($C_{13}$ & $C_{24}$), two leakage capacitors ($C_{12}$ & $C_{34}$) and two cross-coupling capacitors ($C_{14}$ & $C_{23}$).

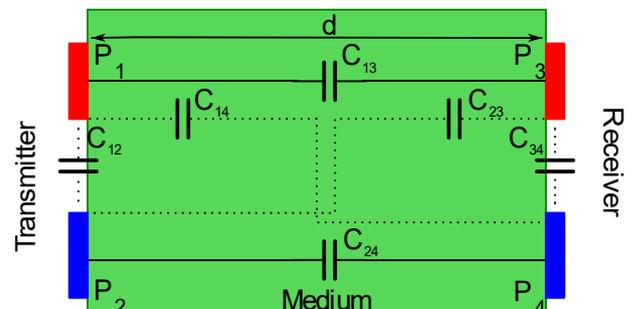
Fig. 3: Four-plate structure with main, leakage and cross-coupling capacitances

The size of these capacitances, especially the main capacitance, is determined by the surface of the plates ($A$), the distance between the plates ($d$) and the dielectric constant of the medium or relative permittivity ($\varepsilon_r$). The main capacitance can be calculated as follows:

$$C = \varepsilon_r \frac{A}{d} \quad (1)$$

The parasitic capacitances ($C_{12}$, $C_{14}$, $C_{23}$, $C_{34}$) are unpractical to determine by analytical calculations. Therefore, the parasitic capacitances can either neglected in analytical calculations or determined by means of finite element simulations. These simulations use a Pi-model as in Fig. 4. This schematic is easier to handle than the one in Fig. 3 [5], [6]. The primary ($C_P$), secondary ($C_S$) and mutual ($C_M$) capacitance than can be calculated as [1], [5], [6]:

$$C_P = \frac{(C_{13} + C_{14}) \cdot (C_{23} + C_{24})}{C_{13} + C_{14} + C_{23} + C_{24}} \quad (2)$$

$$C_S = \frac{(C_{13} + C_{23}) \cdot (C_{14} + C_{24})}{C_{13} + C_{14} + C_{23} + C_{24}} \quad (3)$$

$$C_M = \frac{(C_{13} \cdot C_{24}) - (C_{14} + C_{23})}{C_{13} + C_{14} + C_{23} + C_{24}} \quad (4)$$

In present literature, CPT is often presented with air as the only medium between plates [7]–[9]. This is presumably because CPT has been seen as an alternative to IPT, which already has many application areas where air serves as the medium. However, if another medium increases the energy transfer in CPT, this could open up new application areas where IPT is less viable [10].

Previous studies have shown that the electrical power that can be transferred, depends on the size of the coupling capacitance [2], [8]. In theory, the greater the coupling capacitance, the greater the power that can be transmitted. Considering ( 1 ), the total capacity can be increased by selecting a medium with a large dielectric constant. For this reason, media such as water, insulation material, glass or concrete allow for greater electrical power to be transferred than air [11]–[16]. Their dielectric constants are namely 3 to 80 times greater.

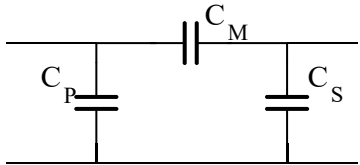

Fig. 4: Pi-model

### III. THEORETICAL ANALYSIS

The capacitances $C_{13}$ & $C_{24}$ of the plate structure in Fig. 3 establish the power transfer. These capacitances are calculated taking the geometry and medium into account using ( 1 ). First, parasitic capacities are neglected here and the capacitive coupling factor $k_e$ is assumed to be one. This means that the primary ($C_P$), secondary ($C_S$) and mutual ($C_M$) capacity in the Pi-model of Fig. 4 equal half the main capacitance [1], [2], [6]. In this paper, mutual capacity $C_M$ is used to compare different mediums as a larger $C_M$ means higher power transfer [8].

For a square plate surface with side $x$ equal to 30 cm, altering the distance from 1 mm to 20 cm, Fig. 5 shows the resulting capacitance $C_M$ for six different materials in the gap. The materials used in this analytical comparison are listed in Table 1 with their dielectric constants.

Table 1: Materials with dielectric constants

| Material | Dielectric constant |
|---|---|
| Air | 1.0058986 [11] |
| Brick | 3.3 [12] |
| Rockwool | 4.7 [13] |
| Concrete | 4.96 [14] |
| Glass | 7.6 [15] |
| Water | 80.103 [16] |

A first finding here is that the resulting $C_M$ with water as a medium is much higher, 80 times according to dielectric constant, than in air. Even typical building materials like glass or rockwool lead to a significantly higher resulting capacitance, 3 to 8 times greater at a distance of 10 mm. In the next section, we will extend the investigation with parasitic capacitances using finite element simulation.

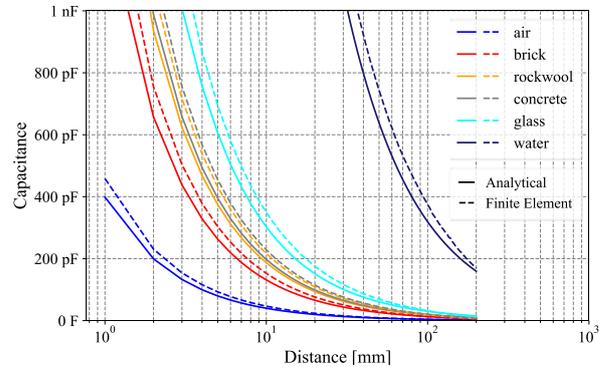

Fig. 5: Influence of medium on analytical mutual capacitance between two plates. The analytical simulations are represented by full lines and finite elements simulations by dashed lines

### IV. SIMULATION RESULTS

#### A. Coupling capacitance

In order to take into account the leakage and cross-coupling capacitances, a finite element simulation software such as CST Studio Suite is required. The simulation model, as in Fig. 6, is based on the same parameters of the analytical calculations in Fig. 3. The model has a symmetric capacitance matrix as output.

$$C = \begin{bmatrix} C_{11} & C_{12} & C_{13} & C_{14} \\ C_{21} & C_{22} & C_{23} & C_{24} \\ C_{31} & C_{32} & C_{33} & C_{34} \\ C_{41} & C_{42} & C_{43} & C_{44} \end{bmatrix} \quad (5)$$

The self-capacitances of the plates ($C_{11}$, $C_{22}$, $C_{33}$, $C_{44}$) are listed at the diagonal of the matrix C in ( 5 ). Those are neglected in this paper because the distance between the plates is much smaller than the distance between the plates and their reference at infinity [17]. Furthermore, the fringing fields, the electrical conductivity of the medium, the equivalent series resistance and the equivalent series inductance of the capacitive coupling are not taken into account in these finite element simulations.

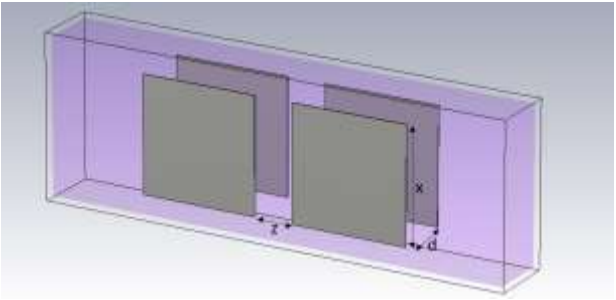

Fig. 6: CST Studio Suite simulation model

For the described CPT system, the resulting mutual capacitances $C_M$ are presented in Fig. 5 by dashed lines. These simulations included the leakage and cross-coupling capacitances.

For a distance up to 25 mm, the simulation results are about 15 % higher than the analytical calculations. This can be explained by the maximum accuracy of the simulation program of one picofarad. After 25 mm, the values become significantly smaller due to the influence of the leakage and cross-coupling capacitance, which can be found in the values of the capacitive coupling coefficient in Fig. 7. Due to the maximum accuracy of the simulation program, the coupling coefficient around a distance of 25 mm, is slightly higher than one.

The capacitive coupling coefficient of water drops faster than the coefficient of air. The advantage of the high permittivity, becomes a disadvantage: cross-coupling and leakage capacities increase and causing the coupling coefficient to drop faster. This reduces the output voltage at the secondary side and deteriorates efficiency of the system [2],[6]. In the next section we will employ the calculated coupling capacitances in an electrical circuit simulation to quantify their effect on the transferrable electrical power.

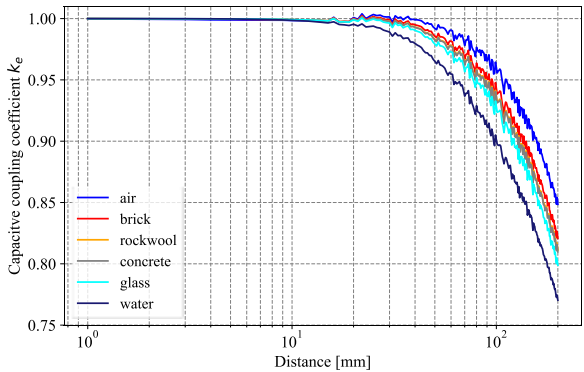

Fig. 7: Influence of a medium on capacitive coupling coefficient $k_e$

### A. Power electronic simulations

In a power electronic simulation, the optimum power transfer is determined for different media. We make use of the software PLECS. The simulated circuit is shown in Fig. 8. A resonator circuit is used with one inductor in series with the coupling capacitors. The latter model the capacitive coupling. This resonator circuit reaches a gain of one when the system is operating at the resonance frequency. Therefore, for each simulation the resonance will be calculated according to ( 6 ) where $L_P$ is the primary inductor and C is the total capacitance of the system. When the calculated frequency becomes greater than 1 MHz, it is assumed that no power transfer is possible due to the actual limits of the power electronic switches that are considered to be MOSFETs [18]. The voltage source feeding the plates has an amplitude of 400 V. The load is simulated with a static resistance of 500 Ω.

$$f_{res} = \frac{1}{2\pi\sqrt{L_P \cdot C}} \qquad (6)$$

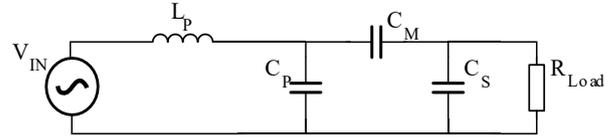

Fig. 8: PLECS simulation model

Simulations are performed with the parameters from Table 2 resulting in output power as a function of the plate distance in Fig. 9. It shows that for a given setting, glass allows CPT to transfer 1 kW over 3 to 4 mm while in air such a power is not even possible at 1 mm.

Table 2: Simulation parameters

| Parameter | Value |
| --- | --- |
| $V_{in}$ | 400 V |
| $L_P$ | 200 μH |
| $R_{Load}$ | 500 Ω |

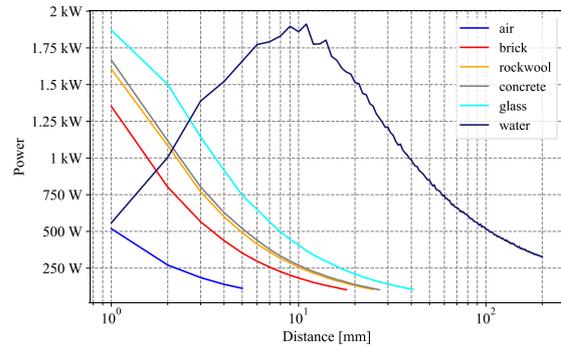

Fig. 9: Influence of a medium on the output power at resonance frequency

According to these simulations, the medium affects the resonant frequency and thus the resulting power output. Media with a high dielectric constant allow higher powers to be transmitted over a greater distance. This is demonstrated by water and glass in Fig. 9. CPT in water and glass each reach a maximum power output of approximately 2 kW while the maximum power output in air is only 500 W. This maximum power is achieved when the impedance of the resonator circuit is at its lowest. For water and glass there is a difference in the distance at which this maximum power was reached, namely for glass this maximum power was reached at 1 mm while in water this maximum power was reached at 11 mm at a lower frequency. The lower frequency makes it possible to transmit power over greater distance, up to 20 cm in water, despite the limitation of a maximum resonant frequency of 1 MHz. An additional benefit of having a lower resonant frequency is that switching losses will be reduced. The resonant frequency for the system in water was throughout all simulations 9 to 12 times smaller than in air.

## Conclusion

During the analytical calculations it was found, as expected, that a medium with a larger dielectric constant than air can have a positive influence on the coupling capacitance. This was further supported by the finite element simulations in which parasitic capacitances were also taken into account. Based on these simulations, the capacitive coupling of capacitive power transfer system with single resonator was simulated. The simulations demonstrated that a medium with a high dielectric constant can ensure higher transmit power over greater distances at lower frequencies. For example, in these simulations glass allows CPT to transfer 1 kW over 3 to 4 mm while in air only 500 W can be reached at 1 mm.

In the future, this research should be extended by taking into account the fringing fields, electrical conductivity of the medium, the equivalent series resistance and the equivalent series inductance of the capacitive coupling. Furthermore, a more detailed modelling of the power converter topology is needed and possible applications must be identified and simulated. Finally, the results should be validated in the lab.